\documentclass[doublespacing]{elsart}

\usepackage{epsfig}

\usepackage{amssymb}

\usepackage{bm}

\usepackage{lineno}

\begin{document}
\begin{frontmatter}

\title{Nonstationary aspects of passive scalar gradient behaviour}

\author{A. Garcia},
\author{M. Gonzalez\corauthref{MG}},
\corauth[MG]{Corresponding author.}
\ead{Michel.Gonzalez@coria.fr}
\author{P. Parantho\"en}
\address{CNRS, UMR 6614, Laboratoire de Thermodynamique, \\ 
CORIA, Site universitaire du Madrillet, \\ 
76801 Saint-Etienne du Rouvray, France}

\begin{abstract}
The dynamics of a passive scalar gradient experiencing
fluctuating velocity gradient through the Lagrangian
variations of strain persistence 
is studied.
To this end,
a systematic, numerical analysis based on the 
equation for the orientation 
of the gradient of a non-diffusive scalar in two-dimensional flow
is performed.
When the gradient responds
weakly
its orientation properties are determined by the mean 
value of strain persistence. Statistical alignment of the 
scalar gradient with the direction defined by the opposed
actions of strain and rotation, by contrast, requires the gradient to
keep up 
with strain persistence fluctuations.
These results 
have been obtained for both strain- and
effective-rotation-dominated regimes and 
are supported by relevant experimental data.
Consequences  
of
the unsteady behaviour of the scalar gradient
on mixing 
properties 
are also analyzed.
\end{abstract}


\end{frontmatter}

\section{Introduction}
\label{sec1}
In random flows, in particular in chaotic or turbulent ones,
the
increase
of the local, instantaneous gradient of a scalar
such as temperature or concentration results in the 
enhancement 
of mixing through acceleration of molecular diffusion.
This ``gradient production'' is caused by the mechanical action of
the velocity gradient, more precisely by 
its straining part, either in the pure straining motion of
hyperbolic regions or in the vicinity of vortices where 
strain stems from differential rotation.
Quite equivalently, at least as far as only convective
mechanisms are considered, 
hastening of the mixing process in random flows
finds expression
in the
stretching of material surface
areas. 
In turbulent flows the growth of the scalar gradient is 
closely connected to the production of small
scales of a scalar field
and to cascade phenomena.
Actually,
the mean dissipation rate of the energy of fluctuations
of a scalar is proportional to the variance of its fluctuating
gradient.

In the view where mixing properties of flows are
explained through the evolution of the 
gradient of a scalar 
the key mechanisms  
rest on the alignment of the gradient with respect to
the local strain principal axes
\cite{P94,S99,Val01,Bal03}.
The rise of the scalar gradient is indeed promoted
by alignment with the most compressive direction of strain.
But
the general problem of scalar gradient alignment is quite 
complicated.
In turbulent flows 
the question amounts to understanding how the gradient
behaves under the combined actions of molecular diffusion
as well as
of
fluctuating
strain and effective rotation 
(i.e. vorticity plus strain basis rotation). 
In three-dimensional turbulence random alignment
of vorticity
and vorticity/strain interaction make the problem even
more complex.

From pure kinematical considerations it is obvious  
that in incompressible flows the scalar gradient is drawn
by the local compressional direction. Yet it is also quite
understandable that the latter is generally not the
equilibrium orientation of the scalar gradient. 
In other words, when not only strain but also rotation
and, possibly, molecular diffusion are present, the fixed
point of the gradient orientation equations is certainly
not the strain compressional axis.
In three-dimensional flows, the bare existence of
this fixed point is even not proved in the general case
\cite{TK94}.
In two-dimensional flows \cite{Lal99} as well 
as in a special, three-dimensional situation \cite{G06,GG06},
however,
the equilibrium orientation and conditions for its existence
have been derived analytically
at least when the scalar is non-diffusive.
It has also been shown that in two-dimensional turbulence
the scalar gradient 
statistically aligns better 
with the local equilibrium direction defined by the balance 
between strain and rotation than with the compressional
direction
\cite{Lal99}.

Even so,
the statistical alignment of the scalar gradient
is determined by the gradient dynamics.
If the velocity field is time varying, alignment with a possible
equilibrium orientation requires that the response 
time scale of the scalar gradient
is
short
enough compared to the 
time scale of 
velocity gradient fluctuations.
Alignment dynamics is thus an essential 
feature of the scalar gradient behaviour and has  
already been addressed in some studies \cite{S99,Bal03,G06,Lal01,Kal00,Gal05}.
The influence on mixing that follows from this 
nonstationary aspect is still not clear.

Trying to understand
how the mixing process is affected by the
dynamics of the scalar gradient is precisely one of
the main goals of the present study.
This is done in Section \ref{sec3} by 
analyzing numerically the growth rate of the gradient
norm for different unsteady regimes
in the two-dimensional case.
We first devote Section \ref{sec2} to the numerical analysis of
alignment statistics when the scalar gradient undergoes
a fluctuating strain persistence. 
This investigation
aims at bearing out the generality of the partial findings of
Ref. \cite{Gal05}. In particular, the orientation equation
of the scalar gradient in a two-dimensional flow
is solved for simulating the strain-dominated as well as the 
effective-rotation-dominated
regimes. 
Different
strain persistence signals are 
used to show the existence 
of a gradient alignment 
that 
is neither the compressional nor the equilibrium direction 
defined by the instantaneous strain persistence.
Section \ref{sec4} reports on experimental results 
supporting the latter numerical study.
Conclusion is drawn in Section \ref{sec5}.
\section{Analysis of statistical scalar gradient alignment}
\label{sec2}
\subsection{Equation for scalar gradient orientation}
\label{sec2.1}
We restrict the analysis to two-dimensional, incompressible flow and
non-diffusive passive scalar. 
Writing the scalar
gradient in the fixed frame of reference as
$ {\bm G} = |{\bm G}|(\cos \theta,\sin \theta) $,
the equation for the gradient orientation 
is \cite{Lal99,Lal01}
\begin{equation}
\label{eq1}
\frac{d \zeta}{d \tau}
=
r - \cos \zeta,
\end{equation}
where
$ \zeta = 2(\theta + \Phi) $ gives the gradient orientation in the 
local strain basis;
angle
$ \Phi $ 
defines the
orientation of the strain principal axes
through
$ \tan \Phi = \sigma_n / \sigma_s $
with
$ \sigma_n = \partial u/\partial x - \partial v/\partial y $
and $ \sigma_s  = \partial v/\partial x + \partial u/\partial y $ 
denoting 
the normal and shear components
of strain, respectively.
Time $ \tau $ is a strain-normalized time
\[
\tau = \int_0^t \sigma(t') \, dt',
\]
with $ t $ standing for the Lagrangian time and $ \sigma $ for
strain intensity, $ \sigma^2 = \sigma_n^2 + \sigma_s^2 $.

Note that Eq. (\ref{eq1}) 
just
proceeds from the usual equation for the passive scalar
(with molecular diffusion neglected) through the equation
for the scalar gradient \cite{Bal03,G06} which is 
handled
as explained in Ref. \cite{Lal99}.

Parameter $ r $ represents the persistence of strain
\cite{Lal99,DT91}
and is defined as
\[
r =
\frac{\omega + 2 d \Phi/d t}{\sigma},
\]
where
$ \omega = \partial v/\partial x - \partial u/\partial y $ is vorticity.
Strain persistence, $ r $, defines an objective criterion
for partitioning the flow; in strain-dominated regions
$ r^2 < 1 $, while in regions where effective rotation
prevails $ r^2 > 1 $.
In  
this article
these regions are occasionally termed hyperbolic and
elliptic respectively.
It is to be noticed that        
this partition 
does not necessarily coincide with
the one derived from the Okubo-Weiss criterion;
because
$ r $ includes the pressure Hessian through
strain basis rotation,
$ d \Phi/d t $,
the corresponding criterion
is more general
\cite{HK98}.

Lapeyre et al. \cite{Lal99}
analyzed Eq. (\ref{eq1})
assuming slow variations of $ r $ along Lagrangian
trajectories and showed the
way in which the
scalar gradient orientation is determined by the local
flow topology. 
For prevailing strain the orientation equation has a 
stable fixed point,
\[
\zeta_{\mbox{\tiny eq}} = - \arccos(r),
\]
which
corresponds to an equilibrium orientation. In the special      
case $ r = 0 $
(i.e. in the pure hyperbolic regime) the equilibrium
orientation, $ \zeta_{\mbox{\tiny eq}} $, coincides with the local
compressional direction,
\[
\zeta_{\mbox{\tiny c}} = -\pi/2.
\]
If effective rotation dominates, 
Eq. (\ref{eq1}) has no fixed point; there is no equilibrium
orientation for the scalar gradient, but a most probable
one given by 
\[
\zeta_{\mbox{\tiny prob}}
= [1 - \mbox{sign}(r)] \pi/2.
\]
Exact balance between strain and effective rotation, namely
$ r^2 =1 $,
drives the scalar gradient to align with the bisector
of strain principal axes.
The general solution, $ \zeta(\tau) $, of Eq. (\ref{eq1})
can be derived for any of the latter three different regimes
\cite{Lal99}.

The study 
of
Garcia et al. \cite{Gal05}  
reveals
that the above approach remains valid
as long as 
the response 
time scale of the scalar gradient is short enough
compared to the time scale of the Lagrangian fluctuations
of $ r $.
They also put forward that
in the opposite case, namely 
when the gradient does 
not keep up with $ r $ fluctuations
and its response is poor,
the alignment of the scalar gradient
is
determined
by the mean value of $ r $, 
$ \langle r \rangle $.
In the following 
we try to generalize 
and support
these results by
the study of
regimes that have not been originally addressed.
Indeed we extend the previous analysis by examining the 
effective-rotation-dominated regime in addition to the strain-dominated one.
Considering
different values of $ \langle r \rangle $
we also confirm the existence of the orientation defined by the mean value
of strain persistence. 

\subsection{Alignment of scalar gradient in the case of fluctuating
strain persistence}
\label{sec2.2}
We simulate 
the fluctuations of strain persistence, $ r $, with
a
stochastic differential equation,
\begin{equation}
\label{eq2}
dr =
- (r - \langle r \rangle) \beta^{\star} d \tau
+
{(2 r'^2 \beta^{\star} d \tau)}^{1/2} \xi.
\end{equation}
In Eq. (\ref{eq2}) $ \xi $ is a standardized Gaussian
random variable.
This equation depends on three parameters, namely
$ \beta^{\star} $, $ \langle r \rangle $ and $ r' $.
Giving $ \beta^{\star} $ a value is equivalent to
prescribe
the nondimensional integral time scale of $ r $, 
$ T^{\star} $, 
through
$ T^{\star} = 1/\beta^{\star} $.
Parameter 
$ r' $ stands for the root mean square of $ r $.
The
few data on strain persistence statistics
\cite{Lal99,Gal05}
show it is not a
Gaussian 
variable.
However, the actual statistics are immaterial here, for 
the concern is mainly in the gradient response
to time-varying strain persistence.

The evolution of the scalar gradient orientation, then,
is derived 
by solving numerically
\begin{equation}
\label{eq3}
d \zeta = [r(\tau) - \cos \zeta] \, d \tau.
\end{equation}
Since $ d \tau = \sigma dt $, 
it is clear 
from Eq. (\ref{eq3})
that
the response time scale of the gradient 
orientation
to the forcing mechanism 
represented by $ r $
is of the order of $ 1/\sigma $.
Parameter $ \beta^{\star} $ gives a measure of
the gradient response compared to forcing stimulation.
Small values of $ \beta^{\star} $
(resp. slow variations of $ r $)
mean that 
the scalar gradient
responds well to
$ r $ fluctuations, while large values 
(resp. fast variations of $ r $) correspond to a 
poor
response
of the scalar gradient.
In the special case for which strain intensity, $ \sigma $, is constant
$ \beta^{\star} = 1/\sigma T $ 
where $ T $ is the integral time scale of $ r $ signal
and thus $ \beta^{\star} $ is the ratio of gradient response time scale
to $ r $ fluctuations time scale.

Garcia et al. \cite{Gal05} have shown that
the approach of
Lapeyre et al. \cite{Lal99} 
(Section \ref{sec2.1})
does not apply to scalar gradient orientation
if
strain persistence fluctuates too rapidly
for the scalar gradient to respond.
They argued that
in this case the most probable 
orientation of the gradient should be determined
by the mean value of $ r $, $ \langle r \rangle $, 
rather than by the instantaneous
one. In particular,
it was put forward that
if $ {\langle r \rangle}^2 < 1 $
(in other words, if the regime is hyperbolic
{\em on an average})
and the time scale of $ r $ fluctuations is shorter
than the gradient response time scale,
the preferential orientation of the gradient should be
defined by
\[
\zeta_{\langle r \rangle} = -\arccos\langle r \rangle.
\]
In reality, the most probable orientation cannot be determined
analytically, but is close to $ \zeta_{\langle r \rangle} $.
This results from the fact that the weaker the gradient response,
the closer $ \zeta $ to $ \zeta_{\langle r \rangle} $.
A simplified proof is given in Appendix \ref{appA}.
The behaviour of the scalar gradient when its response
to $ r $ fluctuations is poor 
is confirmed by the present simulations for  
two different values of $ \langle r \rangle $ 
as well as for small and large values of $ \beta^{\star} $.

\begin{figure}[htpb]
\begin{center}
\scalebox{0.5}{\includegraphics{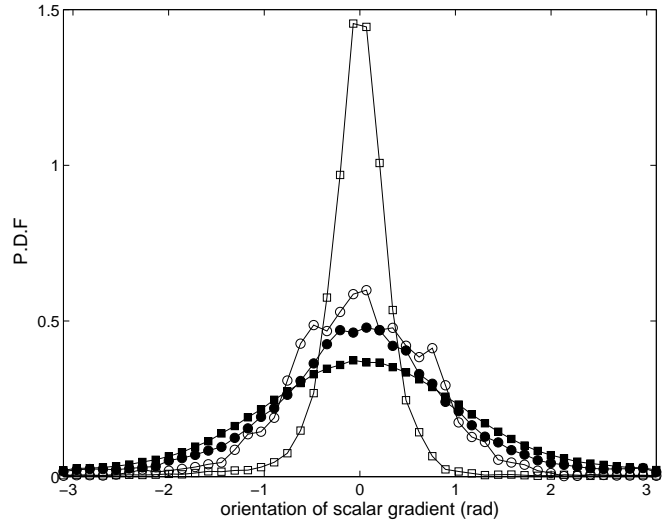}}\\
\caption{\label{fig1}
P.d.f's of scalar gradient orientation for
$ \langle r \rangle = 0 $ and $ r'^2 = 4 $
conditioned on $ r^2 < 1 $
(dominating strain).
P.d.f of $ \zeta - \zeta_{\langle r \rangle}$:
$ \bullet $
$ \beta^{\star} = 5 $,
$ \circ $
$ \beta^{\star} = 0.01 $;
p.d.f of $ \zeta - \zeta_{\mbox{\tiny eq}}$:
$ \blacksquare $
$ \beta^{\star} = 5 $,
$ \square $
$ \beta^{\star} = 0.01 $.}
\end{center}
\end{figure}

The p.d.f's of scalar gradient alignment
have been
derived from the numerical evolution of $ \zeta $ computed 
using Eq. (\ref{eq3})  
and Eq. (\ref{eq2})
with 
$ \langle r \rangle = 0 $ and $ r'^2 = 4 $. 
The fluctuations of
$ r $ thus describe the hyperbolic regime
($ r^2 < 1 $), but also make significant inroads into the elliptic one
($ r^2 > 1 $).
The
p.d.f's of
$ \zeta - \zeta_{\mbox{\tiny eq}} $
and
$ \zeta - \zeta_{\langle r \rangle} $
conditioned on $ r^2 < 1 $ are shown in Fig. \ref{fig1}.
They indicate 
which of the
orientations
$ \zeta_{\mbox{\tiny eq}} $ 
(determined by the instantaneous value of $ r $;
Section \ref{sec2.1})
and
$ \zeta_{\langle r \rangle} $
(defined by $ \langle r \rangle $)
is statistically the best one
in the hyperbolic regime.
Clearly, 
for $ \beta^{\star} = 0.01 $ the scalar gradient 
preferentially
aligns with
the instantaneous, equilibrium direction, 
$ \zeta_{\mbox{\tiny eq}} $,
which is generally different from the compression one,
in agreement with the approach by Lapeyre et al. \cite{Lal99}.
For $ \beta^{\star} = 5 $ it is the alignment with the direction
defined by 
$ \zeta_{\langle r \rangle} $ which is the most probable
one.
Note that for $ \langle r \rangle = 0 $ 
$ \zeta_{\langle r \rangle} = - \pi/2 $
and thus
coincides with the compressional direction.
Interestingly,
in the present case
the scalar gradient thus statistically aligns 
with 
the compressional direction   
because the gradient
response to $ r $ fluctuations is poor ($ \beta^{\star} > 1 $)
and the mean value of $ r $ corresponds to a pure hyperbolic regime
($ \langle r \rangle = 0 $).
This is reminiscent of the experimental situation analyzed
by Garcia et al. \cite{Gal05}. 

\begin{figure}[htpb]
\begin{center}
\scalebox{0.5}{\includegraphics{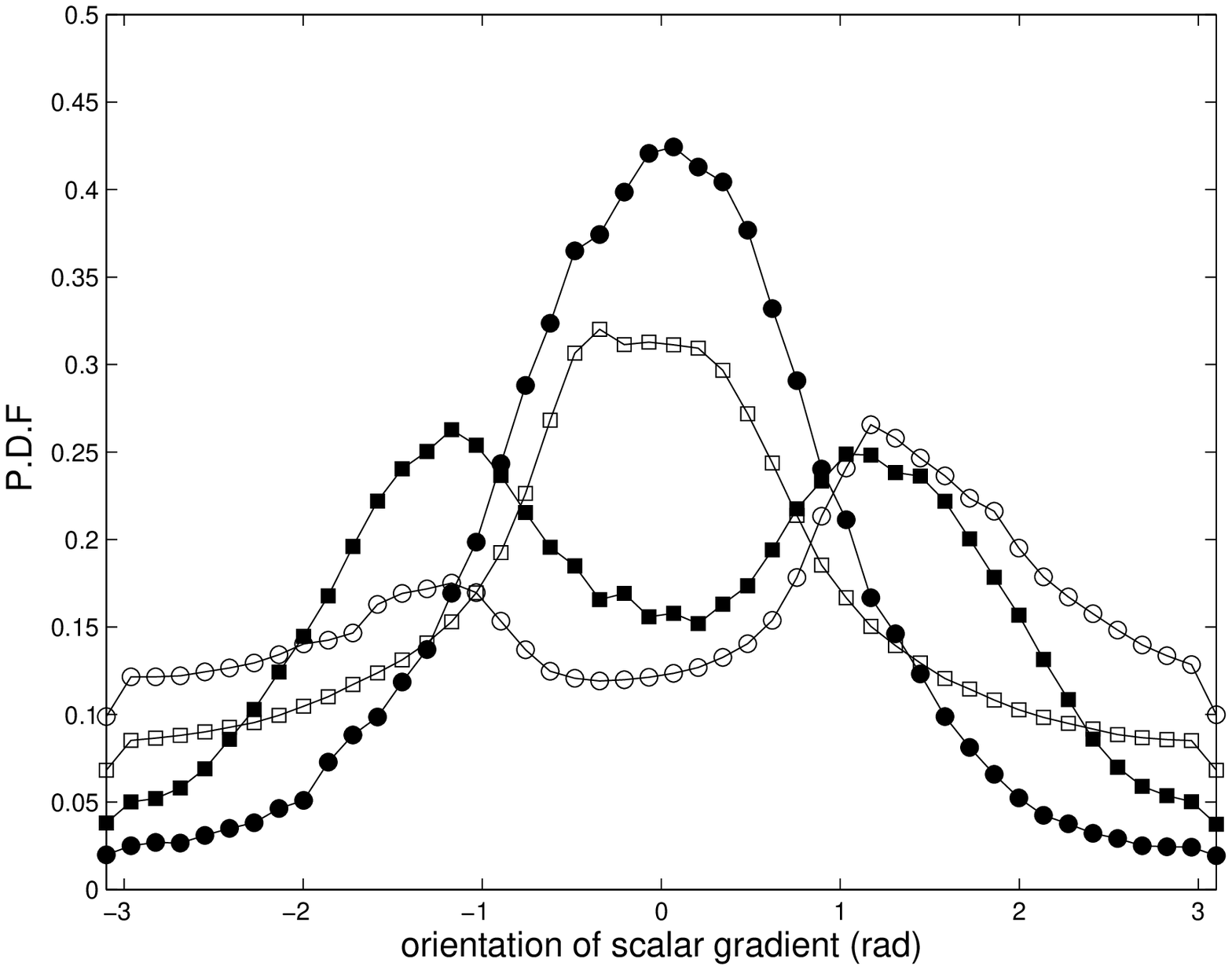}}\\
\caption{\label{fig2}
P.d.f's of scalar gradient orientation for
$ \langle r \rangle = 0 $ and $ r'^2 = 4 $
conditioned on $ r^2 > 1 $
(dominating effective rotation).
P.d.f of $ \zeta - \zeta_{\langle r \rangle}$:
$ \bullet $
$ \beta^{\star} = 5 $,
$ \circ $
$ \beta^{\star} = 0.01 $;
p.d.f of $ \zeta - \zeta_{\mbox{\tiny prob}}$:
$ \blacksquare $
$ \beta^{\star} = 5 $,
$ \square $
$ \beta^{\star} = 0.01 $.}
\end{center}
\end{figure}

The results 
in the elliptic regime 
(p.d.f's conditioned on $ r^2 >1 $; Fig. \ref{fig2}) 
are similar.
For $ \beta^{\star} = 0.01 $ the most probable
gradient orientation 
is given by $ \zeta_{\mbox{\tiny prob}} $ (Section \ref{sec2.1})
which agrees with the analysis of Lapeyre et al. \cite{Lal99}.
In the case $ \beta^{\star} = 5 $ 
the scalar gradient again aligns preferentially with
the direction defined by $ \zeta_{\langle r \rangle} $,
that is, the compressional one.
Yet this is not paradoxical.
When the scalar gradient does not respond to the fluctuations
of $ r $ its orientation is governed by the mean value,
$ \langle r \rangle $, rather than by the instantaneous one 
and remains close to $ \zeta_{\langle r \rangle} $.
This may drive 
the gradient to align with the compressional direction,
even though $ r $ takes elliptic values,
provided that $ \langle r \rangle $ is close to zero.
Note that
the secondary maxima in Fig. \ref{fig2} are explained by
$ |\zeta_{\langle r \rangle} - \zeta_{\mbox{\tiny prob}}|
=
|\zeta_{\mbox{\tiny c}} - \zeta_{\mbox{\tiny prob}}|
= \pi/2 $.

\begin{figure}[htpb]
\begin{center}
\scalebox{0.5}{\includegraphics{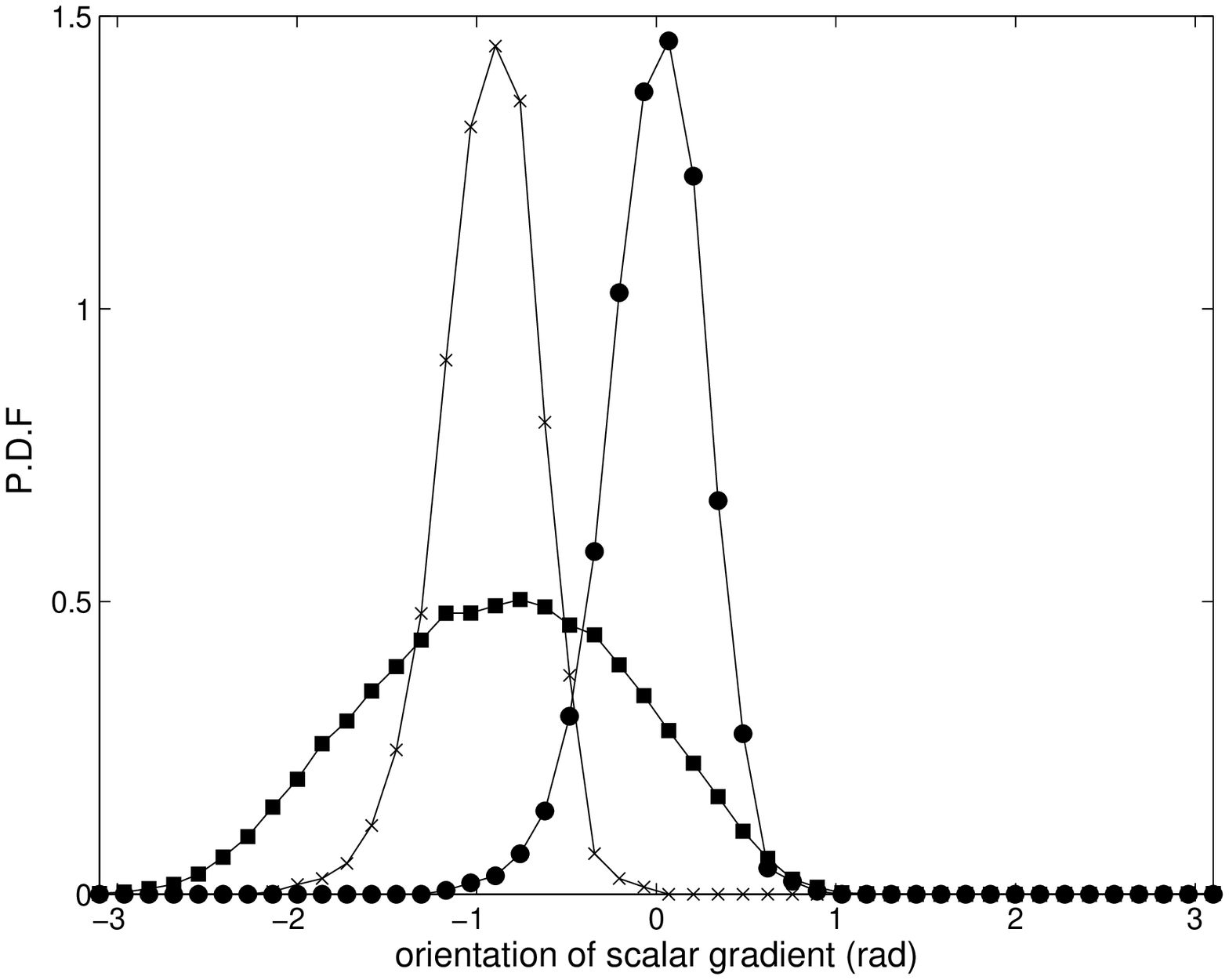}}\\
\caption{\label{fig3}
P.d.f's of scalar gradient orientation for
$ \langle r \rangle = -0.8 $, $ r'^2 = 4 $
and $ \beta^{\star} = 100 $
conditioned on $ r^2 < 1 $ (dominating strain).
$ \bullet $ p.d.f of $ \zeta - \zeta_{\langle r \rangle} $;
$ \blacksquare $ p.d.f of $ \zeta - \zeta_{\mbox{\tiny eq}} $;
$ \times $ p.d.f of $ \zeta - \zeta_{\mbox{\tiny c}} $.}
\end{center}
\end{figure}

\begin{figure}[htpb]
\begin{center}
\scalebox{0.5}{\includegraphics{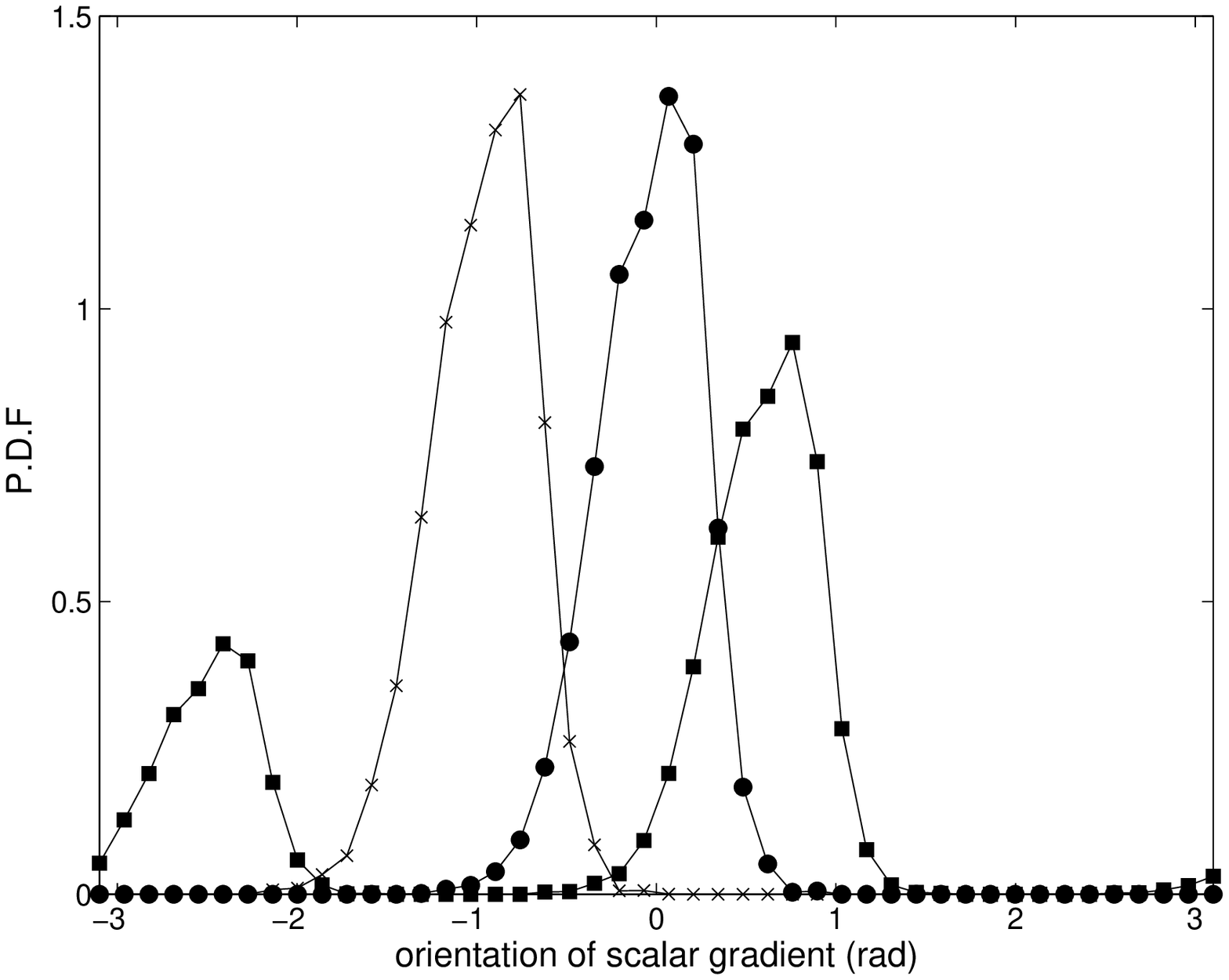}}\\
\caption{\label{fig4}
P.d.f's of scalar gradient orientation for
$ \langle r \rangle = -0.8 $, $ r'^2 = 4 $
and $ \beta^{\star} = 100 $
conditioned on $ r^2 > 1 $ (dominating effective rotation).
$ \bullet $ p.d.f of $ \zeta - \zeta_{\langle r \rangle} $;
$ \blacksquare $ p.d.f of $ \zeta - \zeta_{\mbox{\tiny prob}} $;
$ \times $ p.d.f of $ \zeta - \zeta_{\mbox{\tiny c}} $.}
\end{center}
\end{figure}

Alignment p.d.f's derived with $ \langle r \rangle = -0.8 $
are plotted in Figs. \ref{fig3} and \ref{fig4}.
If
$ \beta^{\star} $ 
is small
the value of $ \langle r \rangle $ is immaterial; 
the scalar gradient preferentially aligns with the equilibrium
direction, $ \zeta_{\mbox{\tiny eq}} $, 
and the same orientation p.d.f's 
as those plotted in Figs. \ref{fig1} and
\ref{fig2} 
for $ \beta^{\star} = 0.01 $ are derived.
Figures \ref{fig3} and \ref{fig4} thus only display the orientation
p.d.f's in the case where 
$ \beta^{\star} $ is given a large value, $ \beta^{\star} = 100 $.
Results for the hyperbolic regime 
are presented in Fig. \ref{fig3}.
Plainly,
$ \zeta_{\langle r \rangle} $ (which is computed as 
$ \zeta_{\langle r \rangle} = -\arccos(-0.8) $)
is
the most probable orientation of the scalar gradient.
Alignment with the equilibrium direction,
$ \zeta_{\mbox{\tiny eq}} $, is weak and
as shown by the p.d.f of $ \zeta - \zeta_{\mbox{\tiny c}} $
there is no trend of the gradient to align with the
compressional direction.

In the elliptic regime (Fig. \ref{fig4}), too, the orientation
defined by $ \langle r \rangle $ is statistically the
best one.
As a matter of course, the gradient does not align with
the compressional direction, but there is also no trend at all toward
the direction defined by $ \zeta_{\mbox{\tiny prob}} $.
\section{``Gradient production'' and mixing properties}
\label{sec3}
With the same assumptions as those stated in Section \ref{sec2.1}
the equation for the norm of the scalar gradient is
\cite{Lal99,Lal01}
\[
\frac{2}{|\bm{G}|} \frac{d |\bm{G}|}{d \tau}
=
-\sin \zeta,
\]
from which it is clear that the mean growth rate of the
gradient norm, $ \rho $, is given by $ \rho = - \langle \sin \zeta \rangle $.
Note that alignment with the compressional direction,
$ \zeta = \zeta_{\mbox{\tiny c}} = - \pi/2 $,
corresponds to the maximum growth rate.

The mean growth rate has been derived from the statistics 
of $ \zeta $ computed by solving Eqs. (\ref{eq2}) and (\ref{eq3}).
The simulations have been 
run
for
three different $ r $ signals, namely
$ r'^2  = 0.1 $, $ r'^2 = 4 $ and $ r'^2 = 16 $
with
$ \beta^{\star} $ ranging from 
0.1 to 100.
For all cases $ \langle r \rangle = 0 $.

\begin{figure}[htpb]
\begin{center}
\scalebox{0.4}{\includegraphics{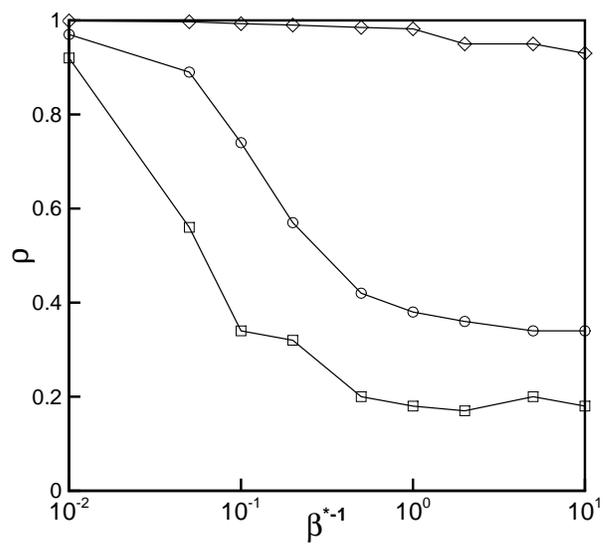}}\\
\caption{\label{fig5}
Mean production rate of scalar gradient norm
{\em vs.} $ {\beta^{\star}}^{-1} $.
$ \diamond $ $ r'^2 = 0.1 $;
$ \circ $ $ r'^2 = 4 $;
$ \square $ $ r'^2 = 16 $.}
\end{center}
\end{figure}

Figure \ref{fig5}
shows the evolution of $ \rho $ {\em vs.} $ {\beta^{\star}}^{-1} $.
Interestingly, $ \rho $ decreases with $ {\beta^{\star}}^{-1} $
whatever the value of $ r'^2 $;
in other words,
the better the response of gradient orientation to 
$ r $ fluctuations (large values of $ {\beta^{\star}}^{-1} $),
the lower the mean growth rate
of the norm.
When it responds to $ r $
fluctuations
the scalar gradient preferentially aligns with 
either $ \zeta_{\mbox{\tiny eq}} $ or $ \zeta_{\mbox{\tiny prob}} $
which are determined by the instantaneous value of $ r $ and
are both mostly different from the compressional direction
(Section \ref{sec2.2}).
Hence the weaker growth rate.
This feature 
gets more pronounced as the variance $ r'^2 $ is increased.
For the lowest variance ($ r'^2 = 0.1 $) $ r $ mostly fluctuates
within the bounds of the hyperbolic regime 
assuming low values
around $ r = 0 $
and
makes $ \zeta_{\mbox{\tiny eq}} $
not much different from the compressional direction. 
Wider fluctuations ($ r'^2 = 4 $ or $ r'^2 = 16 $),
by contrast,
bring about instantaneous orientations $ \zeta_{\mbox{\tiny eq}} $
corresponding to
larger values of $ |r| $
and thus
lying further from the compressional direction; 
in addition,
more frequent,
deep inroads into the elliptic regime
($ r^2 > 1 $) even 
lead the gradient 
to align with $ \zeta_{\mbox{\tiny prob}} $, that is,
$ 45^o $ away from the compressional direction,
during longer time intervals.
As a result, 
for moderate and large values of $ {\beta^{\star}}^{-1} $
the mean growth rate of the scalar gradient norm is lowered
as the amplitude of $ r $ fluctuations is increased.
For the smallest values of $ {\beta^{\star}}^{-1} $, though,
the growth rate is insensitive to $ r $ variance and 
is close to
its maximum value.
Indeed a sluggish response to $ r $ fluctuations
compels
the gradient to remain aligned with the direction defined by
$ \langle r \rangle $ (Section \ref{sec2.2}) which in the present
case 
(for which $ \langle r \rangle = 0 $)
coincides with the compressional direction.

There is an
interesting consequence for mixing properties.
If $ \langle r \rangle \simeq 0 $,
then
favourable
conditions for mixing are not only achieved in the hyperbolic
regime; 
provided that the gradient response
is weak, 
they are also fulfiled
in those cases where large fluctuations of $ r $ span 
both the hyperbolic and elliptic regimes.
More generally, 
when $ r $ fluctuates with $ \langle r \rangle $ remaining
close to 0 it is the poor response of the scalar gradient
to $ r $ fluctuations
that
promotes the best conditions for fast mixing. 
\section{Analysis of experimental data on scalar gradient alignment}
\label{sec4}
As far as 
we know, simultaneous measurements of velocity and scalar 
gradients are scarce.
Apart from experiments in turbulent flows
\cite{Gal04}, 
joint statistics of velocity gradient and temperature
gradient have been measured in a low-Reynolds number,
two-dimensional, B\'enard - von K\'arm\'an street
\cite{G01,Gal02}.
The latter data confirm
that for fast fluctuations of strain persistence
the preferential orientation of the scalar gradient
is determined by the mean strain persistence rather
than by its instantaneous value.

\begin{figure}[htpb]
\begin{center}
\scalebox{2.5}{\includegraphics{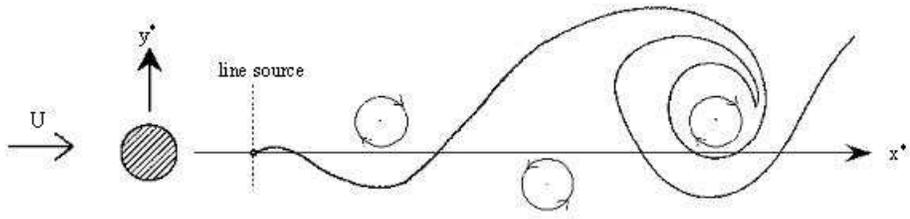}}\\
\caption{\label{fig6}
Experimental set-up (from Godard \cite{G01}).}
\end{center}
\end{figure}

A detailed description of the experiment 
and measurement techniques are given 
in Refs. \cite{G01,Gal02,Pal04}.
In brief, 
the experimental set-up consists of a 
2 mm-diameter ($ D $) circular cylinder used for generating
a two-dimensional B\'enard - von K\'arm\'an street;
the Reynolds number based on the cylinder diameter is
$ \mbox{Re} = 63 $.
Temperature is passively injected through a 20 $ \mu$m-diameter
heated
line source located in the near wake of the cylinder
(Fig. \ref{fig6}).
The line source can be set 
either in or off the centre of the street.
Velocity gradients and temperature gradients 
are
derived from simultaneous measurements of temperature
and velocity components.
These data have been used to  
obtain
Lagrangian statistics of the 
scalar gradient orientation 
conditioned on the strain persistence parameter, $ r $, 
as explained in Refs. \cite{G06,Gal05}.

\begin{figure}[htpb]
\begin{center}
\scalebox{0.5}{\includegraphics{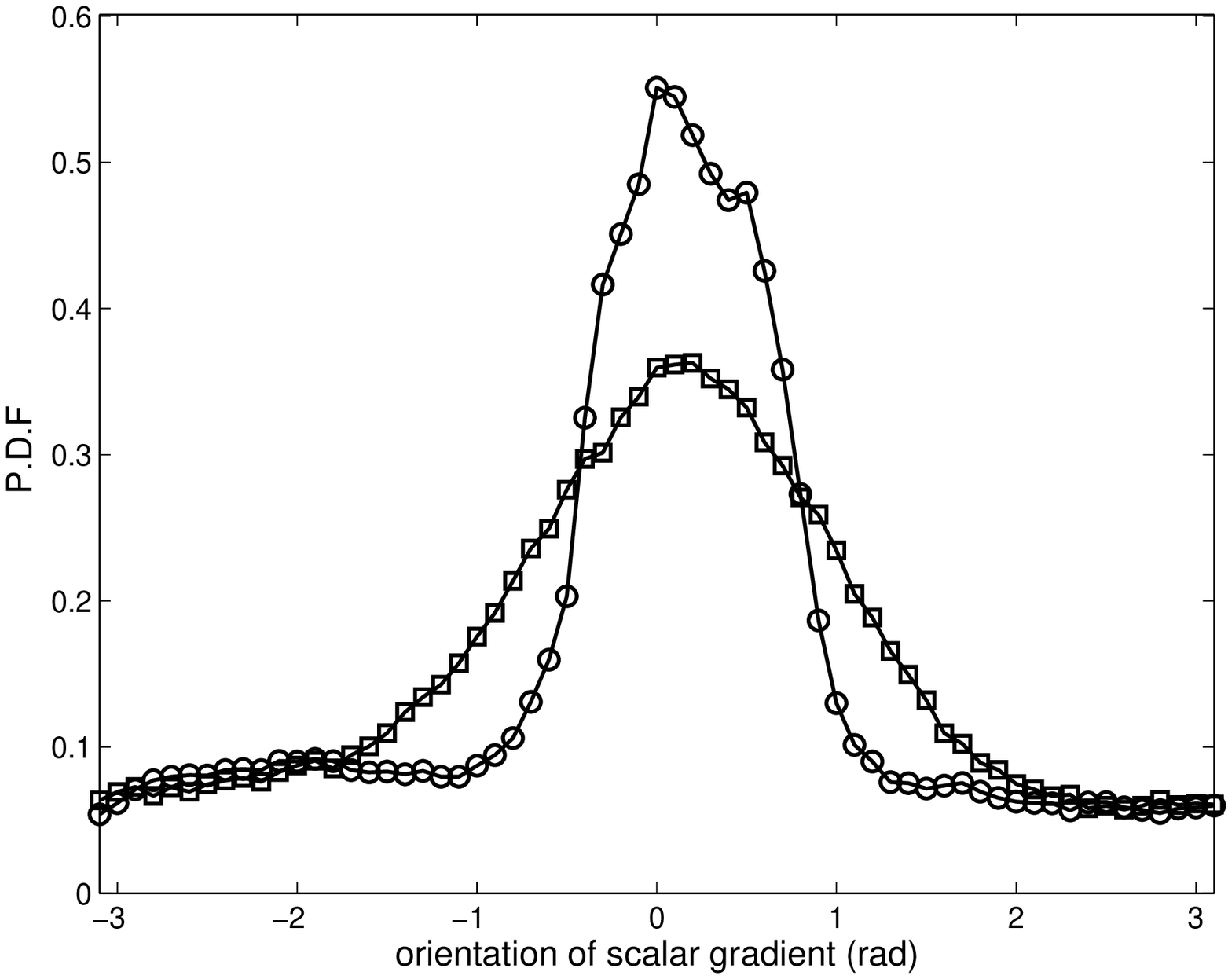}}\\
\caption{\label{fig7}
Experimental
p.d.f's of temperature gradient orientation
conditioned on $ r^2 < 1 $ (dominating strain)
and $ |\bm{G}| > 100 $ in the far field of
the heated line source ($ x/D > 4 $).
The
source is in the centre of the cylinder wake;
experimental $ \langle r \rangle \simeq 0 $.
$ \circ $ p.d.f of $ \zeta - \zeta_{\mbox{\tiny c}} $;
$ \square $ p.d.f of $ \zeta - \zeta_{\mbox{\tiny eq}} $.}
\end{center}
\end{figure}

Part of the latter results have been 
already
analyzed
\cite{G06,Gal05}.
In particular, it has been found that in this experiment 
the temperature gradient does not respond to $ r $
fluctuations. More precisely, 
$ \langle \sigma \rangle T \ll 1 $,
where the mean value of strain, $ \langle \sigma \rangle $,
and $ T $, the autocorrelation time scale of $ r $,
have been computed by averaging over Lagrangian trajectories. 
This previous statistical study was mainly focused on
the hyperbolic zones of the flow,
far enough downstream (i.e. $ (x-x_s)/D > 4 $
with $ x_s $ the distance of the source to the cylinder),
when the
heated
line source is in the centre of the cylinder wake.
The analysis
showed (Fig. \ref{fig7}) that 
the 
alignment of the temperature gradient with the 
compressional direction is statistically better than
with the equilibrium direction derived from the
approach of Lapeyre et al. \cite{Lal99} (Section \ref{sec2.1}).
This result lends support to the findings of the
numerical study presented in Section \ref{sec2.2}.
Most likely,
preferential alignment of the
temperature gradient with the compressional direction 
does not result from any kinematic attraction,
but from the gradient dynamics.
Because of the poor response of the temperature
gradient to $ r $ fluctuations, the preferential
alignment 
is given by $ \langle r \rangle $
computed along Lagrangian trajectories
which, in the above conditions, is found to be close to 0
and thus defines an orientation that almost coincides with
the compressional direction.

Previous studies \cite{Lal01,Cal95} suggest that molecular diffusion
may influence scalar gradient orientation, but 
that
large gradients are weakly affected.
As to the present case,
Garcia et al. \cite{Gal05} have shown that the departure of
the experimental results from the Lapeyre et al. approach
is not to be ascribed to molecular diffusion. 
The latter 
certainly 
plays on the shape of the orientation p.d.f's, but
is not the cause for the compressional direction being more
probable than the equilibrium one.

\begin{figure}[htpb]
\begin{center}
\scalebox{0.5}{\includegraphics{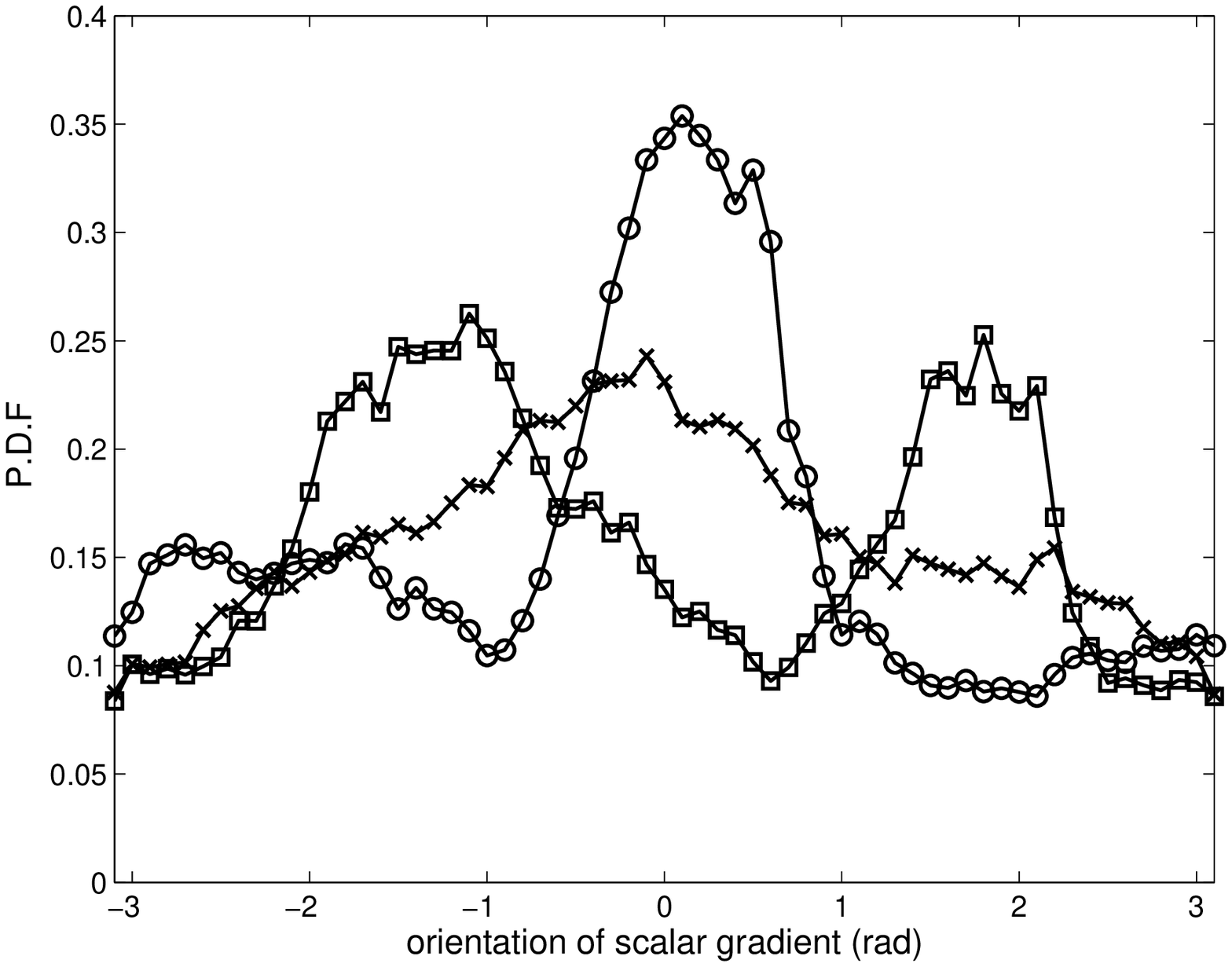}}\\
\caption{\label{fig8}
Experimental
p.d.f's of temperature gradient orientation
conditioned on $ r^2 > 1 $ (dominating effective rotation)
and $ |\bm{G}| > 100 $ in the far field of
the heated line source ($ x/D > 4 $).
The
source is in the centre of the cylinder wake;
experimental $ \langle r \rangle \simeq 0 $.
$ \circ $ p.d.f of $ \zeta - \zeta_{\mbox{\tiny c}} $;
$ \square $ p.d.f of $ \zeta - \zeta_{\mbox{\tiny prob}} $;
$ \times $ p.d.f of $ \zeta - \zeta_{\bm{N}_-} $.} 
\end{center}
\end{figure}

Further
experimental data 
for
the elliptic regions of the flow and in the case where the line
source is located off the centre of the cylinder wake
confirm the above picture. 
Results for elliptic regions when the 
line source is in the centre of the wake are displayed
in Fig. \ref{fig8}.
They concern 
the p.d.f's of orientation of the temperature gradient 
with respect to the compressional direction, 
$ \zeta_{\mbox{\tiny c}} $,
and 
to
direction 
$ \zeta_{\mbox{\tiny prob}} $
imposed by dominating effective rotation
\cite{Lal99} (Section \ref{sec2.1}).
The p.d.f of orientation with respect to direction
$ \zeta_{\bm{N}_-} $ 
is also shown.
The latter direction corresponds to the lowest eigenvalue
of tensor $ \bm{N} $ which is defined as:
\[
\frac{d^2 |\bm{G}|^2}{d t^2}
=
\bm{G}^T \bm{N} \bm{G}
\]
and is related to the pressure Hessian \cite{Kal00}.
According to Klein et al. \cite{Kal00},
when
strain intensity varies significantly along
Lagrangian trajectories
$ \zeta_{\bm{N}_-} $ 
is the preferential orientation of the scalar gradient
in effective-rotation-dominated regions. 
From Fig. \ref{fig8}
it is clear that the temperature gradient does not tend
to align with $ \zeta_{\mbox{\tiny prob}} $.
There is a trend of the gradient to align with
$ \zeta_{\bm{N}_-} $, 
but the best alignment is with the compressional direction.
This striking behaviour 
strongly pleads for the above-described scenario and for the
conclusions of the numerical analysis of Section \ref{sec2.2}:
when strain persistence fluctuations are too fast for the
scalar gradient to respond (which in the experiment is measured
by the low value of $ \langle \sigma \rangle T $)
the latter is blind to the local topology 
and its orientation is entirely governed by the mean value of
the strain persistence parameter.

\begin{figure}[htpb]
\begin{center}
\scalebox{0.5}{\includegraphics{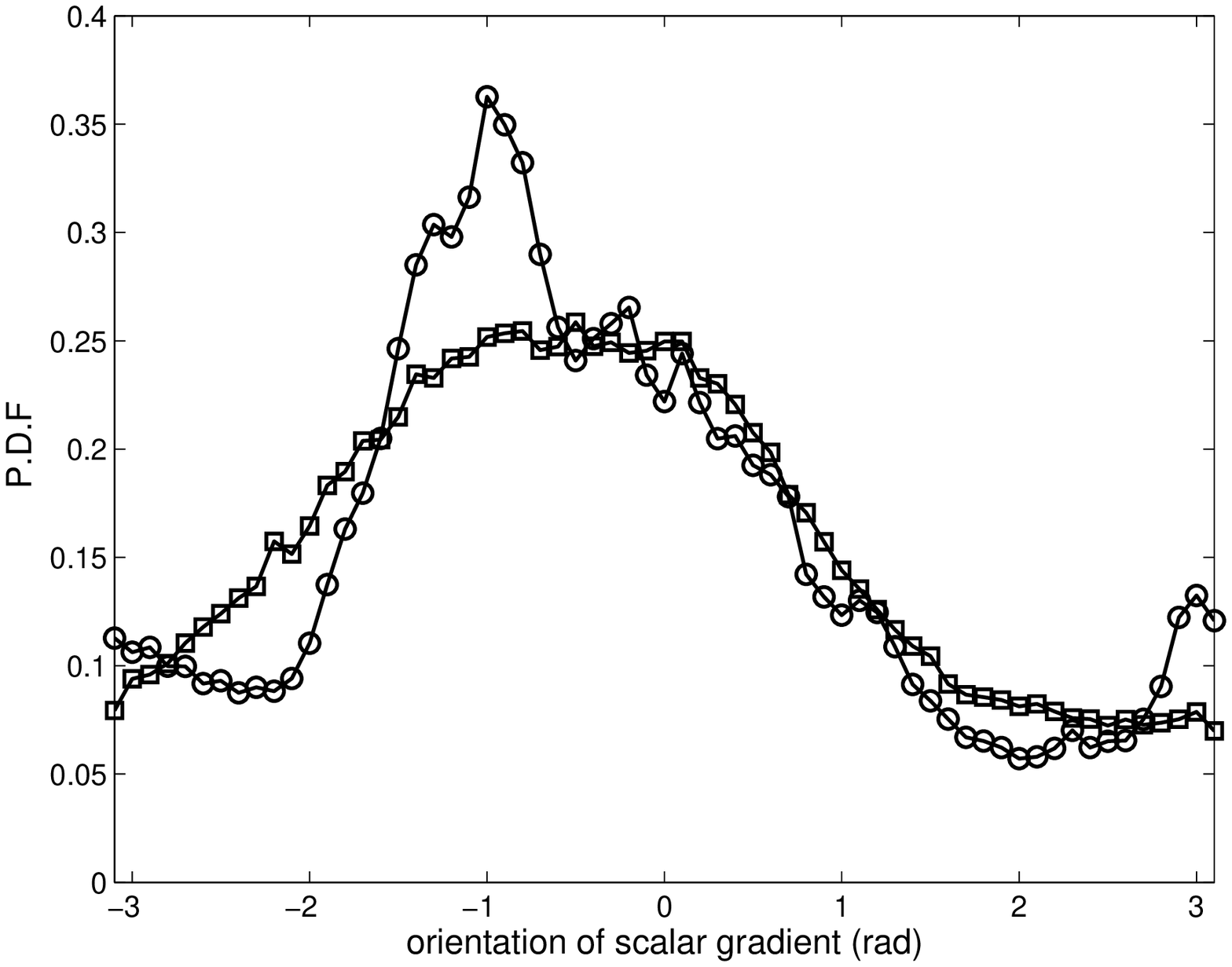}}\\
\caption{\label{fig9}
Experimental
p.d.f's of temperature gradient orientation
conditioned on $ r^2 < 1 $ (dominating strain)
and $ |\bm{G}| > 100 $ in the far field of
the heated line source ($ x/D > 4 $).
The
source is off the centre of the cylinder wake
($ y/D = 0.1 $);
experimental $ \langle r \rangle \simeq -0.8 $.
$ \circ $ p.d.f of $ \zeta - \zeta_{\mbox{\tiny c}} $;
$ \square $ p.d.f of $ \zeta - \zeta_{\mbox{\tiny eq}} $.}
\end{center}
\end{figure}

\begin{figure}[htpb]
\begin{center}
\scalebox{0.5}{\includegraphics{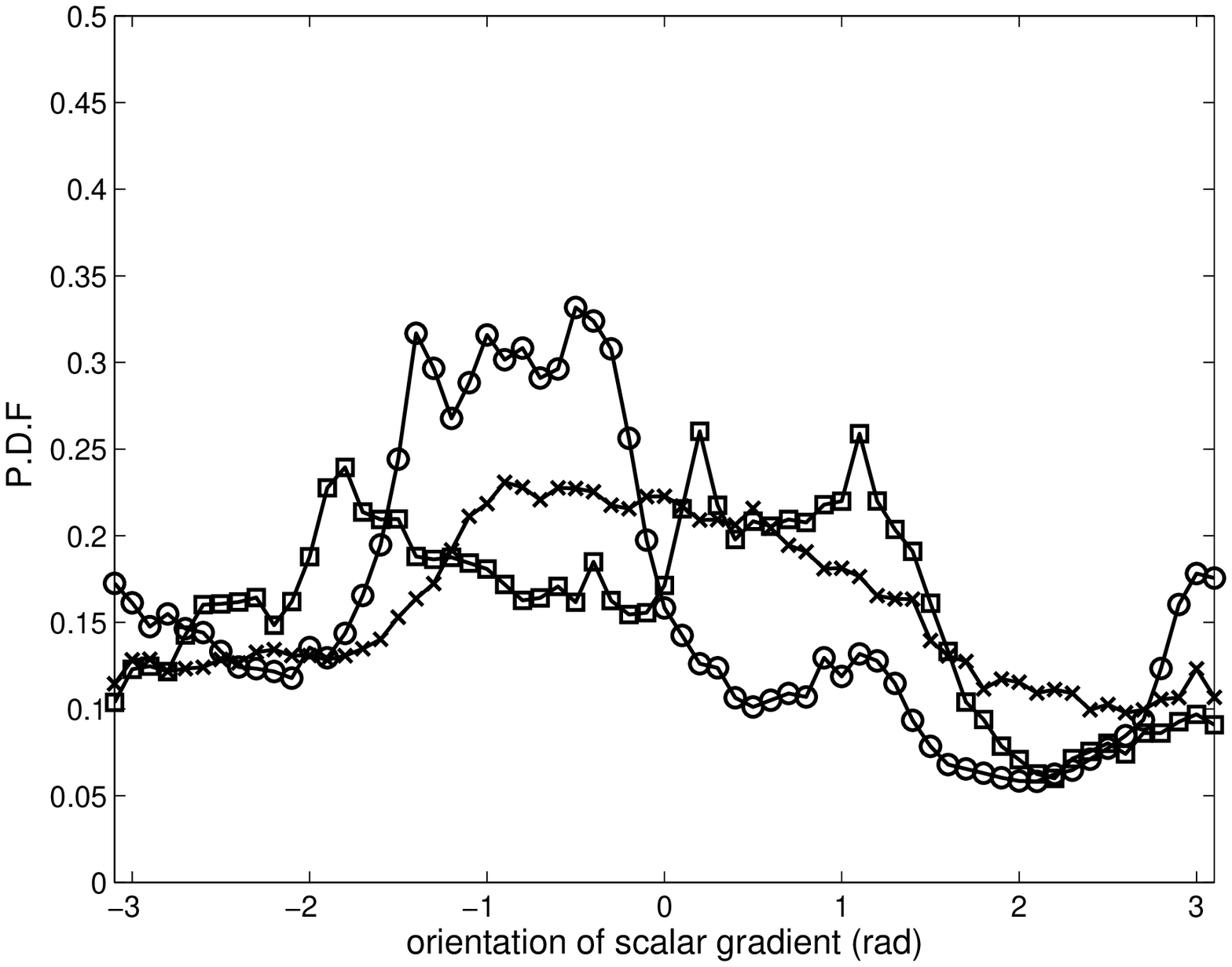}}\\
\caption{\label{fig10}
Experimental
p.d.f's of temperature gradient orientation
conditioned on $ r^2 > 1 $ (dominating effective rotation)
and $ |\bm{G}| > 100 $ in the far field of
the heated line source ($ x/D > 4 $).
The
source is off the centre of the cylinder wake
($ y/D = 0.1 $);
experimental $ \langle r \rangle \simeq -0.8 $.
$ \circ $ p.d.f of $ \zeta - \zeta_{\mbox{\tiny c}} $;
$ \square $ p.d.f of $ \zeta - \zeta_{\mbox{\tiny prob}} $;
$ \times $ p.d.f of $ \zeta - \zeta_{\bm{N}_-} $.} 
\end{center}
\end{figure}

When the heated line source is set off the centre of
the B\'enard - von K\'arm\'an street (more precisely, at
transversal distance
$ y/D = 1 $ from the 
previous position) the mean value $ \langle r \rangle $ of
strain persistence computed for Lagrangian trajectories 
along which
the temperature gradient keeps significant values 
($ |\bm{G}| \geq 100 $)
is
found to be close to -0.8. 
Figure \ref{fig9} shows the p.d.f's of temperature gradient 
orientation with respect to either the compressional 
direction
or the equilibrium
direction, $ \zeta_{\mbox{\tiny eq}} $, derived from the approach
of Lapeyre et al. \cite{Lal99},
in strain-dominated regions.
There is a trend 
of the gradient
to
align with $ \zeta_{\mbox{\tiny eq}} $. However, a much better
statistical alignment is found with the direction corresponding
to $ \zeta - \zeta_{\mbox{\tiny c}} \simeq -1 $, that is, with
$ \zeta \simeq -1-\pi/2 \simeq -\arccos(-0.8) = -\arccos \langle r \rangle $.
P.d.f's of Fig. \ref{fig9} are more distributed and thus
display lower peaks than those corresponding to the source
being in the centre of the cylinder wake
(Fig. \ref{fig7}). This may result from a bigger influence of
rotation on the temperature gradient when the line source
is off the centre of the wake.       
Orientation statistics 
in effective-rotation-dominated regions 
have
similar
trends.
Figure \ref{fig10} shows that the temperature gradient aligns
slightly better with
$ \zeta_{\bm{N}_-} $ 
than with $ \zeta_{\mbox{\tiny prob}} $. 
But
in this case, too, the best statistical alignment
-- broader though the p.d.f is -- is found around
$ \zeta \simeq -\arccos(-0.8) $.
These latter experimental data with $ \langle r \rangle \neq 0 $
thus give a more 
general support to the previous findings    
on
the dynamics of
scalar gradient orientation. 
\section{Conclusion}
\label{sec5}
Analysis
of the behaviour of the passive scalar gradient
undergoing the influence of varying velocity derivatives
through fluctuating strain persistence 
uncovers some interesting results regarding both the
statistical
gradient orientation and mixing properties.

Generalizing the analysis of Garcia 
et al. \cite{Gal05} to the effective-rotation-domina\-ted
regime, we confirm that the statistics of scalar gradient
alignment depend on the gradient response to strain
persistence fluctuations.
More precisely, the numerical study based on the equation for
the scalar gradient orientation in two-dimensional flow
shows that perfect response 
drives the gradient to preferentially
align 
with a direction determined 
by the instantaneous value of strain persistence
as predicted by the analysis of Lapeyre et al. \cite{Lal99}.
When the response of the scalar gradient is poor, though,
the preferential alignment is given by the mean value
of strain persistence
which indicates whether the flow regime is, on an average,  
either strain- or effective-rotation-dominated.
It follows interestingly
that 
the scalar gradient
preferentially aligns with the compressional direction
provided that
it is almost insensitive to
strain persistence fluctuations
and
the mean strain persistence is close
to the pure hyperbolic value.
This result opposes the usual statement that it is
a
perfect response to the fluctuating rotation
of the strain basis (which in the two-dimensional case is explicit
in the strain persistence parameter) that 
leads the gradient to align with the compressional direction.

The general picture derived from the numerical study
is firmly supported by experimental, Lagrangian          
joint statistics of velocity gradient and scalar gradient
derived from simultaneous measurements of velocity
and temperature in a two-dimensional, low-Reynolds
number
B\'enard - von K\'arman street.
It is worth noting that its low Reynolds number  
marks this flow from the simulated two-dimensional turbulent flows
in which
scalar gradient alignment was previously studied
\cite{Lal99,Lal01}.
In particular, it may be that alignment with the equilibrium
direction found by Lapeyre et al. \cite{Lal99} results from
a good response of the scalar gradient to $ r $ fluctuations
in turbulent flows (moderate or large $ \sigma T $). Checking 
this surmise would require Lagrangian data on strain persistence
in two-dimensional turbulence.
Regarding the Lagrangian 
properties of strain persistence and scalar gradient
dynamics,
another interesting and unanswered  issue is whether the
present flow is a special one 
or belongs to a more general class of flows.

Finally,
unsteady behaviour of the
scalar gradient 
may unexpectably
affect
mixing properties 
through alignment statistics.
An interesting result is that
poor
response of the gradient to
strain persistence fluctuations  
does not
inevitably
oppose efficient
mixing.
In particular,
the study
of the growth rate of the gradient norm
clearly shows that
when the mean strain persistence is close to
the pure hyperbolic value 
(in other words, strain is persistent on an average)
it is the
weak
response of the gradient
to strain persistence fluctuations
which promotes
the highest growth rate and thus the best conditions for
mixing. 
\begin{appendix}
\section{Behaviour of scalar gradient orientation
\label{appA}
for fast fluctuations of strain persistence}
We restrict to the case where $ {\langle r \rangle}^2 < 1 $
and $ \zeta_{\langle r \rangle} = -\arccos \langle r \rangle $
is defined.

From Eq. (\ref{eq1}) the equation for the difference
$ \zeta^{\star} = \zeta - \zeta_{\langle r \rangle} $ is:
\begin{equation}
\label{eqA1}
\frac{d \zeta^{\star}}{d \tau}
=
r - \langle r \rangle \cos \zeta^{\star} - \gamma \sin \zeta^{\star}
\end{equation}
with $ \tau = \int_0^t \sigma(t') dt' $ and
$ \gamma = {(1 - {\langle r \rangle}^2)}^{1/2} $.

We get rid of the nonlinearity of Eq. (\ref{eqA1}) by
assuming small values of $ \zeta^{\star} $. 
To first order, then:
\begin{equation}
\label{eqA2}
\frac{d \zeta^{\star}}{d \tau}
+
\gamma \zeta^{\star} = r'
\end{equation}
with $ r' = r - \langle r \rangle $.
The proof is thus not general, but is given as an 
additional
argument for the behaviour of the scalar gradient
when its response to $ r $ fluctuations is poor.

We assume a sine signal for $ r' $,
$ r' = a \sin \omega^{\star} \tau $,
in which frequency 
is normalized as $ \omega^{\star} = \omega/\langle \sigma \rangle $
with $ \langle \sigma \rangle = \tau/t $.
The nondimensional
frequency, $ \omega^{\star} $, 
thus gives a measure of the response 
of the scalar gradient to $ r $ fluctuations.
In particular, poor response is expected for large values
of $ \omega^{\star} $.

The solution of Eq. (\ref{eqA2}) is:
\begin{equation}
\label{eqA3}
\zeta^{\star}
=
\left[
\zeta^{\star}(0) + \frac{a \omega^{\star}}{{\omega^{\star}}^2
                                           + \gamma^2}
\right]
\exp(-\gamma \tau)
+ \frac{a}{\gamma(1 + {\omega^{\star}}^2/\gamma^2)}
\left(\sin \omega^* \tau - \frac{\omega^{\star}}{\gamma} 
                      \cos \omega^{\star} \tau \right).
\end{equation}

Note that Eq. (\ref{eqA3})
can be recast in the more standard form:
\[
\zeta^{\star}
=
\left[
\zeta^{\star}(0) + \frac{a \omega^{\star}}{{\omega^{\star}}^2
                                           + \gamma^2}
\right]
\exp(-\gamma \tau)
+ \frac{a}{\gamma{(1+{\omega^{\star}}^2/\gamma^2)}^{1/2}}
  \sin(\omega^{\star} \tau - \phi)
\]
with $ \cos \phi = {(1 + {\omega^{\star}}^2/\gamma^2)}^{-1/2} $
and phase, $ \phi $, is present when the response to the
stimulation is not perfect.

From Eq. (\ref{eqA3})
the long-time evolution of $ \zeta^{\star} $ is given by:
\[
\zeta^{\star}
\simeq
\frac{a}{\gamma(1 + {\omega^{\star}}^2/\gamma^2)}
\left(\sin \omega^* \tau - \frac{\omega^{\star}}{\gamma} 
                      \cos \omega^{\star} \tau \right).
\]
For $ \omega^{\star} \gg 1 $:
\[
\zeta^{\star}
\simeq
\frac{a}{{\omega^{\star}}^2/\gamma}
\left(\sin \omega^* \tau - \frac{\omega^{\star}}{\gamma} 
                      \cos \omega^{\star} \tau \right).
\]
Since $ \omega^{\star}/\gamma \gg 1 $, 
$ \zeta^{\star} $ is bounded as:
\[
-\frac{a}{\omega^{\star}} \leq \zeta^{\star} \leq \frac{a}{\omega^{\star}}.
\]

It follows that
when $ r $ fluctuates on a much shorter time scale
than the response time scale of the scalar gradient
so much so that the latter does not keep up with
$ r $ fluctuations the gradient orientation, $ \zeta $,
remains close to $ \zeta_{\langle r \rangle} $.
Increasing the amplitude of $ r $
fluctuations, however, leads to the opposite behaviour.
\end{appendix}

\newpage
\centerline{FIGURE CAPTION}

\bigskip

\noindent
FIG. 1
P.d.f's of scalar gradient orientation for
$ \langle r \rangle = 0 $ and $ r'^2 = 4 $
conditioned on $ r^2 < 1 $
(dominating strain).
P.d.f of $ \zeta - \zeta_{\langle r \rangle}$:
$ \bullet $
$ \beta^{\star} = 5 $,
$ \circ $
$ \beta^{\star} = 0.01 $;
p.d.f of $ \zeta - \zeta_{\mbox{\tiny eq}}$:
$ \blacksquare $
$ \beta^{\star} = 5 $,
$ \square $
$ \beta^{\star} = 0.01 $.

\bigskip

\noindent
FIG. 2
P.d.f's of scalar gradient orientation for
$ \langle r \rangle = 0 $ and $ r'^2 = 4 $
conditioned on $ r^2 > 1 $
(dominating effective rotation).
P.d.f of $ \zeta - \zeta_{\langle r \rangle}$:
$ \bullet $
$ \beta^{\star} = 5 $,
$ \circ $
$ \beta^{\star} = 0.01 $;
p.d.f of $ \zeta - \zeta_{\mbox{\tiny prob}}$:
$ \blacksquare $
$ \beta^{\star} = 5 $,
$ \square $
$ \beta^{\star} = 0.01 $.

\bigskip

\noindent
FIG. 3
P.d.f's of scalar gradient orientation for
$ \langle r \rangle = -0.8 $, $ r'^2 = 4 $
and $ \beta^{\star} = 100 $
conditioned on $ r^2 < 1 $ (dominating strain).
$ \bullet $ p.d.f of $ \zeta - \zeta_{\langle r \rangle} $;
$ \blacksquare $ p.d.f of $ \zeta - \zeta_{\mbox{\tiny eq}} $;
$ \times $ p.d.f of $ \zeta - \zeta_{\mbox{\tiny c}} $.

\bigskip

\noindent
FIG. 4
P.d.f's of scalar gradient orientation for
$ \langle r \rangle = -0.8 $, $ r'^2 = 4 $
and $ \beta^{\star} = 100 $
conditioned on $ r^2 > 1 $ (dominating effective rotation).
$ \bullet $ p.d.f of $ \zeta - \zeta_{\langle r \rangle} $;
$ \blacksquare $ p.d.f of $ \zeta - \zeta_{\mbox{\tiny prob}} $;
$ \times $ p.d.f of $ \zeta - \zeta_{\mbox{\tiny c}} $.

\bigskip

\noindent
FIG. 5
Mean production rate of scalar gradient norm
{\em vs.} $ {\beta^{\star}}^{-1} $.
$ \diamond $ $ r'^2 = 0.1 $;
$ \circ $ $ r'^2 = 4 $;
$ \square $ $ r'^2 = 16 $.

\bigskip

\noindent
FIG. 6
Experimental set-up (from Godard \cite{G01}).

\bigskip

\noindent
FIG. 7
Experimental
p.d.f's of temperature gradient orientation
conditioned on $ r^2 < 1 $ (dominating strain)
and $ |\bm{G}| > 100 $ in the far field of
the heated line source ($ x/D > 4 $).
The
source is in the centre of the cylinder wake;
experimental $ \langle r \rangle \simeq 0 $.
$ \circ $ p.d.f of $ \zeta - \zeta_{\mbox{\tiny c}} $;
$ \square $ p.d.f of $ \zeta - \zeta_{\mbox{\tiny eq}} $.

\bigskip

\noindent
FIG. 8
Experimental
p.d.f's of temperature gradient orientation
conditioned on $ r^2 > 1 $ (dominating effective rotation)
and $ |\bm{G}| > 100 $ in the far field of
the heated line source ($ x/D > 4 $).
The
source is in the centre of the cylinder wake;
experimental $ \langle r \rangle \simeq 0 $.
$ \circ $ p.d.f of $ \zeta - \zeta_{\mbox{\tiny c}} $;
$ \square $ p.d.f of $ \zeta - \zeta_{\mbox{\tiny prob}} $;
$ \times $ p.d.f of $ \zeta - \zeta_{\bm{N}_-} $. 

\bigskip

\noindent
FIG. 9
Experimental
p.d.f's of temperature gradient orientation
conditioned on $ r^2 < 1 $ (dominating strain)
and $ |\bm{G}| > 100 $ in the far field of
the heated line source ($ x/D > 4 $).
The
source is off the centre of the cylinder wake
($ y/D = 0.1 $);
experimental $ \langle r \rangle \simeq -0.8 $.
$ \circ $ p.d.f of $ \zeta - \zeta_{\mbox{\tiny c}} $;
$ \square $ p.d.f of $ \zeta - \zeta_{\mbox{\tiny eq}} $.

\bigskip

\noindent
FIG. 10
Experimental
p.d.f's of temperature gradient orientation
conditioned on $ r^2 > 1 $ (dominating effective rotation)
and $ |\bm{G}| > 100 $ in the far field of
the heated line source ($ x/D > 4 $).
The
source is off the centre of the cylinder wake
($ y/D = 0.1 $);
experimental $ \langle r \rangle \simeq -0.8 $.
$ \circ $ p.d.f of $ \zeta - \zeta_{\mbox{\tiny c}} $;
$ \square $ p.d.f of $ \zeta - \zeta_{\mbox{\tiny prob}} $;
$ \times $ p.d.f of $ \zeta - \zeta_{\bm{N}_-} $. 

\newpage

\newpage

\newpage

\newpage

\newpage

\newpage

\newpage

\newpage

\newpage

\newpage

\end{document}